\documentclass[a4paper,12pt]{article}
\usepackage{graphicx,psfrag,color}
\usepackage{amssymb}
\usepackage{epstopdf}
\def\etap{\eta^{\prime}}
\def\Br{\it Br}%\mbox{\ensuremath{\cal B}

\def\corr#1{\textcolor{black}{#1}}
\def\corrn#1{\textcolor{black}{#1}}

\begin{document}
%\maketitle
\renewcommand{\thefootnote}{\fnsymbol{footnote}}
 \rightline{LAL16-011}
\vspace{.3cm} 
{\Large
\begin{center}
{\bf Novel approach to measure the leptonic $\eta^{(\prime)}\to \mu^+\mu^-$ decays via charmed meson decays}
\end{center}}
\vspace{.3cm}

\begin{center}
N. T. Huong$^a$, E. Kou$^b$ and B. Viaud$^b$\\
\vspace{.6cm}\small

\emph{$^a$ 
Faculty of Physics, VNU University of Science, 
    Vietnam National University} \\
    \emph{ 334 Nguyen Trai, Thanh Xuan, Hanoi, Vietnam}
    
 \emph{$^b$ 
Laboratoire de l'Acc\'el\'erateur Lin\'eaire,
  Univ. Paris-Sud, CNRS/IN2P3 (UMR 8607)} \\
\emph{Universit\'e Paris-Saclay, 91898 Orsay  C\'{e}dex, France}
\end{center}
%\today

\vspace{.3cm}
\hrule \vskip 0.3cm
\begin{center}
\small{\bf Abstract}
\end{center}
In this \corr{article}, we propose a novel approach to measure the branching ratios of the leptonic $\eta^{(\prime)}\to \mu^+\mu^-$ decays by using charmed meson decays, namely,  {$D^{+}_{(s)}\rightarrow \pi^+ \eta^{(\prime)} (\rightarrow \mu^+\mu^-)$} and {$D^{0}\rightarrow K^-\pi^+ \eta^{(\prime)} (\rightarrow \mu^+\mu^-)$}. We \corr{advocate} that the data available at LHCb can already yield a new measurement of $Br(\eta\rightarrow \mu^+\mu^-)$ \corr{with accuracy competitive with} the current world average. We also estimate that using the data collected  by LHCb between 2015
and 2018 in proton-proton collisions at a centre-of-mass energy of 13 TeV, corresponding to an integrated luminosity of 5.0 fb$^{-1}$, the relative uncertainties to this branching ratio can be reduced down to $\sim 10$\%. We also show that the first observation of $Br(\eta^{\prime}\rightarrow \mu^+\mu^-)$ may be possible \corr{with} the Upgrade of the LHCb experiment. 

\vskip 0.3cm \hrule \vskip 1.2cm

\section{Introduction}
\begin{figure}
\begin{center}
\begin{minipage}{5cm}
\includegraphics[width=5cm]{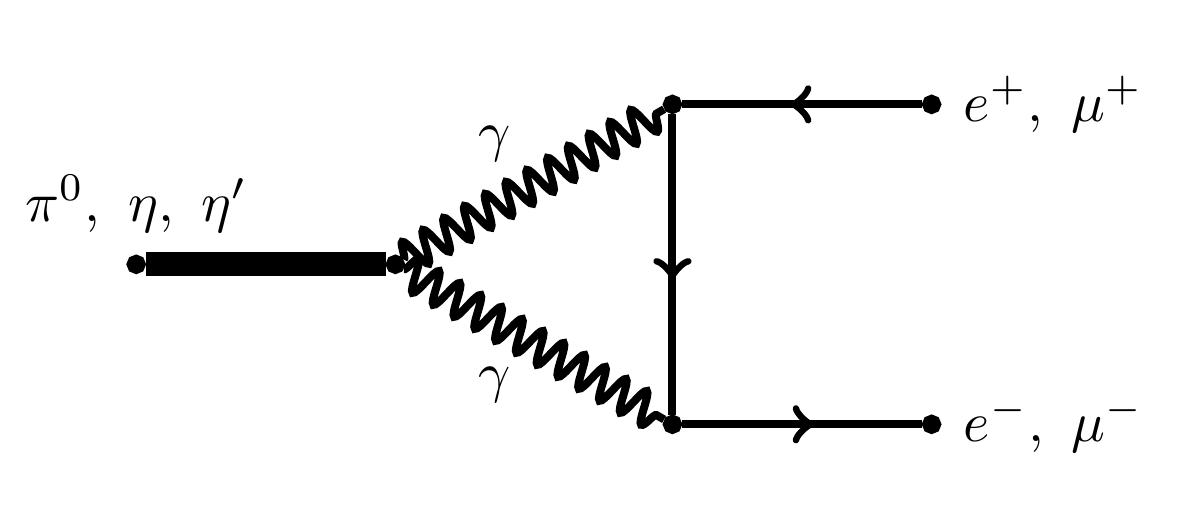}\\
\hspace*{2cm}(a)
\end{minipage}
\begin{minipage}{4cm}
\includegraphics[width=4cm]{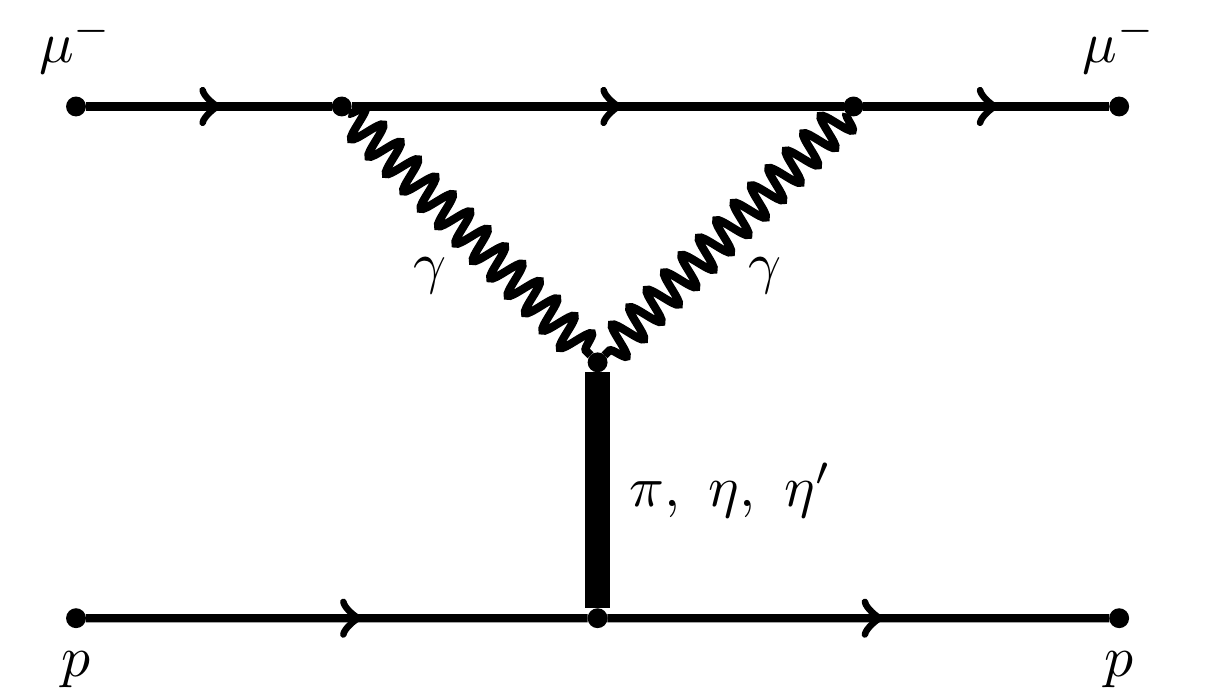}\\
\hspace*{2cm}(b)
\end{minipage}
\begin{minipage}{4.5cm}
\includegraphics[width=4.5cm]{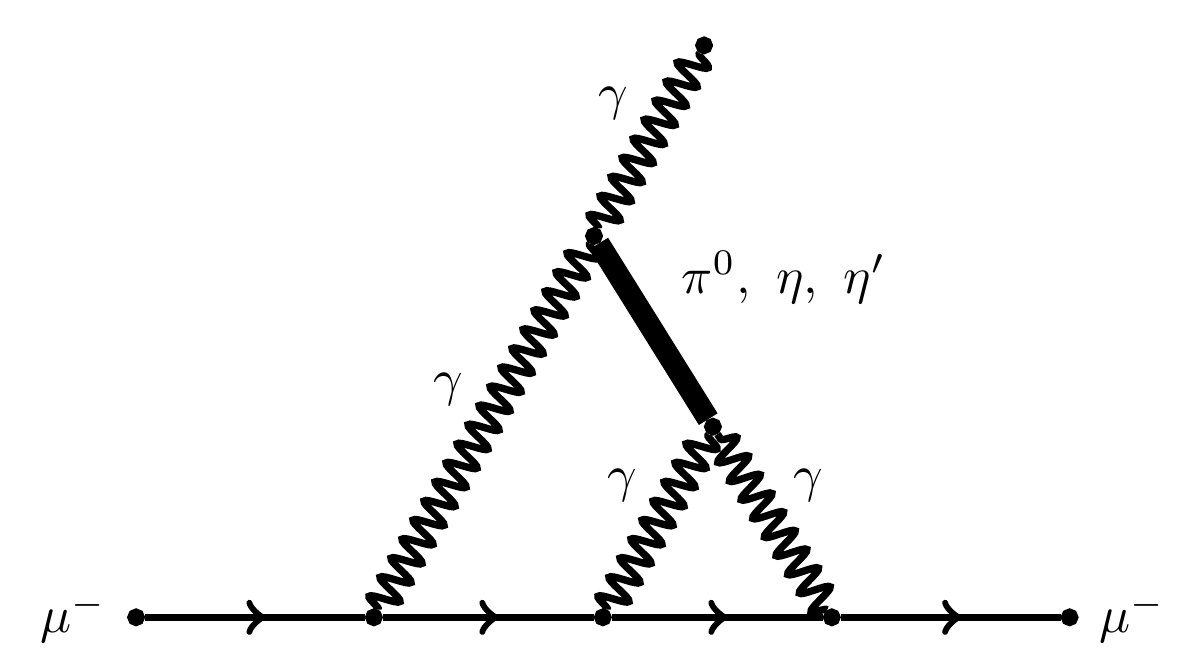}\\
\hspace*{2cm}(c)
\end{minipage}
\caption{Feyman    diagrams for (a)    the leptonic decay of
pseudoscalar mesons \corr{and examples of various phenomena which contain this coupling, (b) the pion contributions to the hyperfine splitting of the muonic hydrogen,} (c) the  \corrn{dominant}  light-by-light contribution
to the muon anomalous magnetic moment.}
\label{Fig:1}
\end{center}
\end{figure}

In this \corr{article}, we discuss \corr{a possible improved measurement of} the rare  $\pi^0\to l^+l^-$ and  $\eta^{(\prime)}\to l^+l^-$ ($l=e, \mu$) decays. The pseudoscalar mesons $\pi^0, \eta, \etap$, being $J^{PC}=0^{-+}$, decay into a two-photon final state while they can also decay into two leptons. \corr{The leptonic decays are very} rare because of being loop induced processes as shown in Fig.~\ref{Fig:1} (a). \corr{This loop diagrams are ultraviolet divergent, which requires a local counter term. These contributions are parameterized by the chiral coupling constants, $\chi_1$ and $\chi_2$~\cite{Savage:1992ac}, and they have to be determined by e.g. using the experimental measurements of the branching ratios of the leptonic decays of  $\pi^0, \eta, \etap$.}

\corr{The determination of the $\chi_{1,2}$ coupling constants has important consequences in various phenomenological studies. Let us give some examples. }

\corr{Recently, two of the authors have computed the pion exchange contributions to the hyperfine splitting of the muonic hydrogen~\cite{Huong:2015naj}. This contribution which comes from the diagram given in Fig.~\ref{Fig:1} (b) contains the same coupling as Fig.~\ref{Fig:1} (a). It was found that the theoretical uncertainty for the pion exchange contribution to the  hyperfine splitting is indeed dominated by the uncertainties in the $\chi_{1,2}$ parameters, thus, it can be improved by measurements of $Br(\pi^0(\eta) \to l^+l^-)$ with higher accuracy.} 

\corrn{An important motivation is the} theoretical prediction of the muon $g-2$ anomalous magnetic moment. The muon $g-2$ is one of the most precisely measured observable in particle physics: \corr{$a_\mu^{\rm exp}=0.00116592089(54)(33)$.}
The theoretical prediction within the Standard Model (SM), $a_\mu^{\rm SM}$  is also very precise while the comparison shows 3.1 $\sigma$ deviation, according to the latest review~\cite{Nyffeler:2016gnb}: 
\corr{$a_\mu^{\rm exp}-a_{\mu}^{\rm SM}=(27.8\pm 8.8)\times 10^{-10}$.} As the experimental uncertainties will be reduced by future experiments namely at Fermilab and J-PARC~\cite{Gorringe:2015cma},  the dominant uncertainty will be theoretical one, in particular, the so-called hadronic vacuum polarization effect and the light-by-light contributions. 
\corrn{The dominant contributions to the hadronic light-by-light amplitude from pseudoscalar meson exchange are as shown in Fig.~\ref{Fig:1} (c)}. This contribution  can be described by two counter-terms in the chiral perturbation theory~\cite{RamseyMusolf:2002cy, Knecht:1999gb} and \corr{one of them is the $\chi_{1,2} $ coupling constants.  The importance of determining these parameters from the leptonic decays has been also} re-emphasized recently in~\cite{Nyffeler:2016gnb, Masjuan:2015cjl}, since in particular, a tension between theory and experiment has been reported in the case of $\pi^0\to e^+e^-$~\cite{Abouzaid:2006kk}.

\corrn{In this article}, we propose to measure the $\chi_{1,2} $ coupling constants by using the semi-leptonic decays of charmed mesons, \corr{in particular, \ensuremath{\Br}$(D^{+}_{(s)}\rightarrow \pi^+ \eta^{(')}(\rightarrow \mu^+\mu^-) )$ and \ensuremath{\Br}$(D^{0}\rightarrow K^-\pi^+ \eta^{(')}(\rightarrow \mu^+\mu^-) )$. These charm decays are recently studied in detail at LHCb~\cite{D2PiMuMu, Aaij:2015hva} for  searching  new physics signals (see~\cite{Cappiello:2012vg,deBoer:2015boa}, for examples of new physics scenarios). }

In section 2, we first introduce the counter-term in terms of chiral Lagrangian and constrain them by using available experimental data and compare to \corr{the theoretical model computation by the lowest meson dominant model (LMD)}, which is often applied to compute the hadronic light-by-light contributions of muon $g-2$. Then, we propose a measurement of $\eta^{(\prime)}\to \mu^+\mu^-$ at LHCb and \corr{assess} the future sensitivity in section 3 and we conclude in section 4. 

%%%%%%%%%%%%%%%%%%%%%%%%%%%%%%%%%%%%%%
\section{Determining the counter-term of leptonic pseudoscalar decays}
In this section, we derive ratios of decay rates: 
\begin{equation}
R_{P}=\frac{\Gamma(P\to l^+l^-)}{\Gamma (P\to \gamma\gamma)}, 
\end{equation}
where $P=\pi^0, \eta, \etap$. 
The $P\to \gamma\gamma$ decay widths are: 
\begin{equation}
\Gamma (\pi^0\to \gamma\gamma)=\frac{m_\pi^3\alpha^2}{64 \pi^3 F_{\corr{\pi}}^2}, \quad 
\Gamma (\eta\to \gamma\gamma)=\frac{m_\eta^3\alpha^2}{192 \pi^3 F_{\corr{\pi}}^2}, \quad 
\Gamma (\etap\to \gamma\gamma)=\frac{m_{\etap}^3\alpha^2}{24 \pi^3 F_{\corr{\pi}}^2}
\end{equation}
which comes from the Wess-Zumino term: 
\begin{eqnarray}
\mathcal{L}^{\rm WZ}&=&\frac{\alpha N_C}{4\pi}Tr( Q^2 \frac{\pi^i \lambda_i}{F_{\corr{\pi}}}) F_{\mu\nu}\tilde{F}^{\mu\nu} \nonumber \\
&=& \frac{\alpha}{4\pi F_{\corr{\pi}}}\left(\pi^3+\frac{\pi^8}{\sqrt{3}}+\frac{2\sqrt{2}\pi^0}{\sqrt{3}}\right)\corr{F_{\mu\nu}\tilde{F}^{\mu\nu}}
\end{eqnarray}
where $Q=(2/3, -1/3, -1/3)_{\rm diag.}$ is the electric charge of the $u, d$ and  $s$ quarks and where  $\lambda_i$ are the Gell-mann matrices extended to the nonet symmetry with $\lambda_0=\sqrt{2/3}(1,1,1)_{\rm diag.}
$ and where $\tilde{F}_{\mu\nu}=1/2\epsilon^{\mu\nu\alpha\beta}F_{\alpha\beta}$. For simplicity, we identify $\eta_8 (\eta_0)$ to $\eta (\etap)$ for now. As we will see later-on, the different factors for $\pi, \eta, \etap$ cancel in the ratio $R_P$ so that in fact,  we do not need to consider the pseudoscalar mixing angle as long as we use this ratio. 

The loop diagram in Fig. 1 (a) has been computed in~\cite{Knecht:1999gb, Quigg:1968zz}. The real part diverges and it requires a local counter term, which can be written to the lowest order in the chiral expansion, as~\cite{RamseyMusolf:2002cy}: 
\begin{equation}
{\mathcal{L}}_{\rm c.t.} = \frac{3i\alpha^2}{32\pi^2}\overline{l}\gamma^\mu\gamma_5 l\left[\chi_1 Tr(Q^2 \Sigma^\dagger \partial_\mu \Sigma-Q^2 \Sigma\partial_\mu \Sigma^\dagger)+\chi_2Tr(Q\Sigma^\dagger Q \partial_\mu \Sigma - Q \Sigma Q \partial_\mu\Sigma^\dagger)\right], 
\end{equation}
where $\Sigma=e^{i\pi^i\lambda_i/F_{\corr{\pi}}}$. The coefficients $\chi_{1,2}$ are renormalization scale  and scheme dependent. Using this counter term, we can write the decay rates as: 
\begin{eqnarray}
&\Gamma (\pi^0\to l^+l^-)=\frac{m_l^2m_\pi\alpha^4}{32 \pi^5 F_{\corr{\pi}}^2}\sqrt{1-\frac{4m_l^2}{m_{\pi}^2}}|A(m_\pi^2)|^{\corr{2}}, \quad 
\Gamma (\eta\to l^+l^-)=\frac{m_l^2m_\eta\alpha^4}{96 \pi^5 F_{\corr{\pi}}^2}\sqrt{1-\frac{4m_l^2}{m_{\eta}^2}}|A(m_\eta^2)|^{\corr{2}}, &\nonumber \\
&\Gamma (\etap\to l^+l^-)=\frac{m_l^2m_{\etap}\alpha^4}{12 \pi^5 F_{\corr{\pi}}^2}\sqrt{1-\frac{4m_l^2}{m_{\etap}^2}}|A(m_{\etap}^2)|^{\corr{2}}, &
\end{eqnarray} 
where the amplitude is given with a loop function as~\cite{Knecht:1999gb}:
\begin{equation}
A(s)=-\frac{\chi_1(\mu)+\chi_2(\mu)}{4}+\frac{N_C}{3}\left[-\frac{5}{\corr{2}}+\frac{3}{2}\ln \left(\frac{m_l^2}{\mu^2}\right)+C(s)\right], 
\end{equation} 
where 
\begin{equation}
C(s)=\frac{1}{\beta_l(s)}\left[{\rm Li}_2\left(\frac{\beta_l(s)-1}{\beta_l(s)+1}\right)+\frac{\pi^2}{3}+\frac{1}{4}\ln^2\left(\frac{\beta_l(s)-1}{\beta_l(s)+1}\right)\right], 
\end{equation}
and $\beta_l(s)=\sqrt{1-\frac{4m_l^2}{s}}$. 
\corr{This result implies} that $R_P$ ($P=\pi, \eta, \etap$) can be expressed in a simple form as: 
\begin{equation}
R_P=\frac{2m_l^2\alpha^2}{m_P^2\pi^2}\corr{\beta_l(m_P^2)}|A(m_P^2)|^{\corr{2}}. 
\end{equation}
As mentioned earlier, the overall factor cancels.

\begin{table}[htdp]
\begin{center}
\begin{tabular}{|c||c|c|c|}
\hline 
$P$& $\pi$ & $\eta$ & $\etap$ 
 \\ \hline \hline
$Br(P\to \gamma \gamma)^{\rm exp}$ & $(9.8823 \pm 0.0034)\times 10^{-1}$& $(3.841\pm 0.020)\times 10^{-1}$ & $(2.2\pm 0.08)\times 10^{-2}$  \\ \hline \hline
$Br(P\to \mu^+\mu^-)^{\rm exp}$ & --- & $(5.8\pm 0.8)\times 10^{-6}$ & $\cdots$  \\ \hline 
$R_P (\mu^+\mu^-)^{\rm exp}$ & --- & $(1.47\pm 0.20)\times 10^{-5}$ &  $\cdots$ \\ \hline 
$R_P (\mu^+\mu^-)^{\rm LMD}$ & --- & $(1.5\pm0.2)\times 10^{-5}$ &  $(6.5\pm 0.1)\times 10^{-6}$ \\ \hline \hline
$Br(P\to e^+e^-)^{\rm exp}$ & %$(6.46\pm 0.33)\times 10^{-8}$
$(7.48\pm 0.38)\times 10^{-8}$ & $<5.6 \times 10^{-6}$ & $<2.1 \times 10^{-7}$  \\ \hline 
$R_P  (e^+e^-)^{\rm exp}$ & %$(6.52\pm 0.334)\times 10^{-8}$ 
$(6.96\pm 0.36)\times 10^{-8}$
&$<1.42 \times 10^{-5}$ & $< 9.5\times 10^{-6}$  \\ \hline 
$R_P (e^+e^-)^{\rm LMD}$ & $(6.2\pm 0.3)\times 10^{-8}$ & $(1.1\pm 0.5)\times 10^{-8}$ &  $(5.5\pm 0.2)\times 10^{-9}$ \\ \hline 
\end{tabular}
\end{center}
\caption{Measured branching ratios taken from PDG~\cite{Agashe:2014kda} and the extracted $R_P$ values and their LMD model prediction using the counter term values given in Eq.~(\ref{eqv0:14}). The \corr{ratio} $R_P(e^+e^-)$ is obtained by using the range $s_{e^+e^-}>0.95 m^2_\pi$.}
\label{Tab:1}
\end{table}%

In~\cite{Knecht:1999gb}, the counter term is computed using the Lowest Meson Dominant model (LMD) as:  
\begin{equation} 
\chi_1(m_\rho)+\chi_2(m_\rho)=-8.8\pm 3.6
\label{eqv0:14}
\end{equation}
where systematic theoretical error of 40 \% is assumed. Using this value, we compute $R_P$ for all the channels. The results are given in Table~\ref{Tab:1}.  We can see that  
the theoretical prediction for $\eta\to \mu^+\mu^-$ shows a good agreement with the LMD prediction while in the $\pi\to e^+e^-$ channels a small deviation appears. Note that for $Br(\pi^0\to e^+e^-)$,  we use the latest KTeV measurement of ~\cite{Abouzaid:2006kk} which subtracts the radiative $\pi^0\to e^+e^-\gamma$ process. 

\begin{figure}
\begin{center}
\includegraphics[width=10cm]{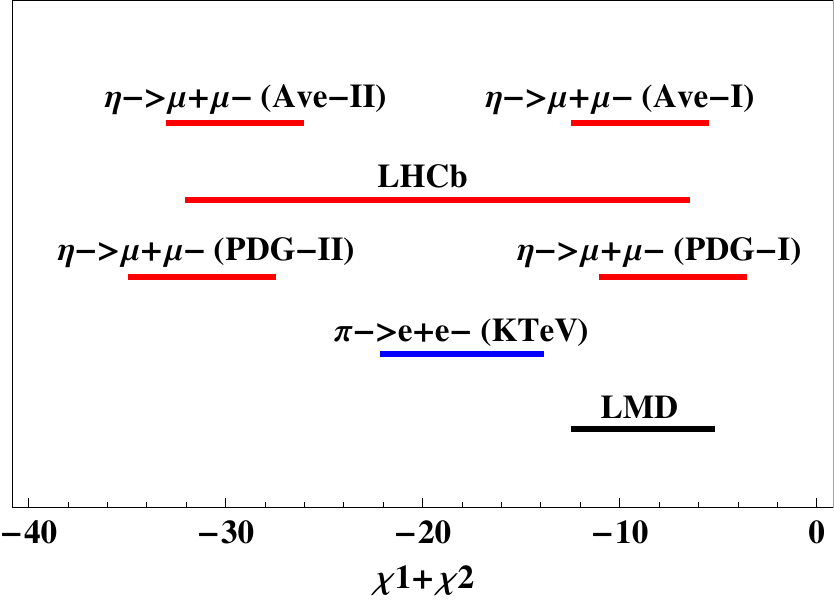}
\caption{Current constraints on the parameter $\chi_1(m_\rho)+\chi_2(m_\rho)$ from the KTeV  experiment for $\pi\to e^+e^-$ (blue) as well as the theoretical prediction by the Lowest Meson Dominant model (black). The current world average  measurement $Br(\eta\to \mu^+\mu^-)=(5.8\pm 0.8)\times 10^{-6}$~\cite{Agashe:2014kda}  leads to two solutions, marked as ``PDG-I, II''. The result using our average based on LHCb's data in~\cite{D2PiMuMu}, $Br(\eta\to \mu^+\mu^-)=(4.7\pm 1.0)\times 10^{-6}$,  is also plotted: the reason for the large error is explained more in detail in the text (also see Fig.~\ref{Fig:3}). The average of the PDG value and LHCb value is given as ``Ave-I, II''. }
\label{Fig:2}
\end{center}
\end{figure}

The counter term, in turn, can be extracted from the experimental values \corr{up to two hold ambiguity}, namely those of $\eta\to \mu^+\mu^-$ and $\pi\to e^+e^-$ (including NLO QED corrections as in~\cite{Vasko:2011pi, Husek:2014tna}): 
\begin{eqnarray}
\left(\chi_1(m_\rho)+\chi_2(m_\rho)\right)_{\eta\to \mu^+\mu^-}^{\rm exp}&=&-6.8\pm 3.6 \quad {\rm or} \quad -32\pm 3.6.  \label{eqv0:15} \\
%\left(\chi_1(m_\rho)+\chi_2(m_\rho)\right)_{\pi \to e^+e^-}^{\rm exp}&=&-24\pm 4 
\left(\chi_1(m_\rho)+\chi_2(m_\rho)\right)_{\pi \to e^+e^-}^{\rm exp}&=&-18\pm 3.9 \quad {\rm or}  \quad 78 \pm 3.9,  \end{eqnarray}
The current situation is summarized in Fig.~\ref{Fig:2}. We drop the second solution of $\pi\to e^+e^-$  for now as it is far from any other measurement. 
As we can see, the experimental extraction of the $\chi_1+\chi_2$ parameter has much larger errors than the theoretical prediction based on LMD model: if we choose the first solution for $\eta \to \mu^+\mu^-$, the average with $\pi\to e^+e^-$ leads to the range of: 
\begin{equation}
\left(\chi_1(m_\rho)+\chi_2(m_\rho)\right)_{\rm average}^{\rm exp}=-12.4\pm 6.0
\label{Eq:15_v1}
\end{equation}
Further improvements in determining the $\chi_{1,2}$ parameters through the experimental measurement of the leptonic pseudoscalar decays are desired. In the next section, we present a novel way to extract this parameter via charmed meson decays. 

%%%%%%%%%%%%%%%%%%%%%%%%%%%%%%%%%%%
\section{Measuring the $\eta^{(\prime)}\to \mu^+\mu^-$ decays via charmed meson decays at LHCb}
\label{sec:LHCbSensitivity}
The LHC produces a huge number of charmed hadrons, like the $D^+
$, $D^+_s$ or $D^0$ mesons, whose decays can be studied by the LHCb
collaboration.
While investigating $D_{(s)}^+ \to \pi^+ \mu^+\mu^-$ decays in the
data collected in 2011~\cite{D2PiMuMu}, LHCb found a few \corrn{dozens of} events
peaking around the $\eta$ meson mass in the dimuon mass spectrum. These events
seem to come from the high branching \corr{ratio} channels  $D_{(s)}^+ \to
\pi^+\eta$, followed by the rare decay $\eta\to \mu^+\mu^-$.
The branching \corr{ratio} {\ensuremath{\Br}$(\eta\rightarrow
\mu^+\mu^-)$} evaluated with this data has a larger uncertainty
than the present world average: {\ensuremath{\Br}$(\eta\rightarrow
\mu^+\mu^-)$}=(5.8$\pm$0.8)$\times$10$^{-6}$~\cite{Agashe:2014kda}. While
LHCb accumulates more and more data, \corr{the collaboration} will be able to measure
these $\eta$ peaks with much higher precision. In this section, we
evaluate LHCb's potential for the  measurement of $\eta\to
\mu^+\mu^- $ and its impact on the determination of  the $\chi_1+\chi_2$
parameter.

We also investigate a possible observation of the $\eta^{\prime}\to \mu^+\mu^-$ decay at LHCb. As shown in Table 1, if we treat $\eta^{\prime}$ in the same way as $\pi$ and $\eta$, we expect its \corr{leptonic} branching ratio to be ${\mathcal{O}}(10^{-7})$.   However, as $\eta^{\prime}$ is much heavier than the other mesons, it is quite possible that the mass correction is sizable. Therefore, a measurement of this branching ratio is very important  to understand the applicability of our  theoretical framework.

 \subsection{Sensitivity of the LHCb experiment to {\bf\boldmath {\ensuremath{\Br}$(\eta\rightarrow \mu^+\mu^-)$} }}
\label{sec:LHCbSensitivityEta}
The most precise measurement of \corr{the $\eta\to \mu^+\mu^-$ decay should be obtained by measuring}
{\ensuremath{\Br}$(D^{+}\rightarrow \pi^+ \eta(\rightarrow \mu^+\mu^-) )$}. 
In data collected in proton-proton collisions at a centre-of-mass energy of 7 TeV and corresponding to an integrated luminosity of 1.0
fb$^{-1}$, LHCb \corr{observed} (29$\pm$7) $D^{+}\rightarrow \pi^+ \eta(\rightarrow \mu^+\mu^-)$ decays  in the region of the dimuon mass
spectrum surrounding the
mass of the $\eta$ meson~\cite{D2PiMuMu}. 
An evaluation  of the
corresponding branching fraction made in~\cite{D2PiMuMu} yields:
{\ensuremath{\Br}$(D^{+}\rightarrow \pi^+ \eta(\rightarrow \mu^+\mu^-) )$}=(2.2$\pm$0.6)$\times$10$^{-8}$. 
\corr{The measurement of} this branching
fraction was not the goal of the \corr{analysis}
described in~\cite{D2PiMuMu} and a more detailled analysis is necessary to measure it. 
\corr{On the other hand, by combining the presently available estimate} with {\ensuremath{\Br}$(D^{+}\rightarrow \pi^+ \eta )$}=(3.66$\pm$0.22)$\times$10$^{-3}$~\cite{Agashe:2014kda}, we can obtain \corr{a reasonable precision}: 
{\ensuremath{\Br}$(\eta\rightarrow \mu^+\mu^-)$}=(6.0$\pm$1.7)$\times$10$^{-6}$. 

\begin{figure}
\begin{center}
\includegraphics[width=12cm]{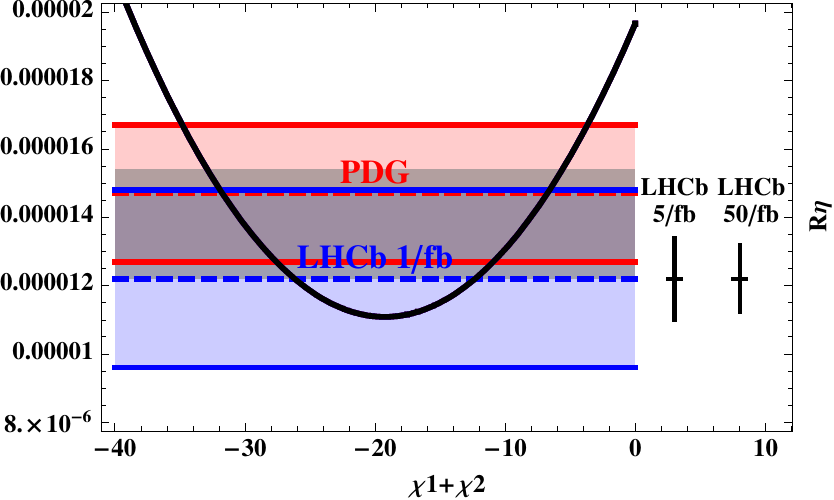}
\caption{The ratio of branching ratios $R_P(\eta\to \mu^+\mu^-)$ is plotted in terms of the parameter $\chi_1(m_\rho)+\chi_2(m_\rho)$. The red bound is from PDG average~\cite{Agashe:2014kda}  and the blue bound is  our average based on LHCb's data in~\cite{D2PiMuMu}. The average of these two measurements is plotted as the gray bound. The relative errors in the branching ratio measurement is 14 \% for PDG and 20 \% for LHCb. In the future the  error obtained at the LHCb can be reduced down to 10 \% level (Run II, $5fb^{-1}$ of data). To reduce this error further at the Upgrade of LHCb ($50fb^{-1}$) improvement in the measurement of the experimental reference as well as normalization channels become necessary (see text for the details). Without this improvement, the error would be limited at 8 \% level as shown on this plot. }
\label{Fig:3}
\end{center}
\end{figure}

The decay 
$D^{+}_{s}\rightarrow \pi^+ \eta(\rightarrow \mu^+\mu^-)$ can also be used. Its branching fraction is also evaluated in 
Ref.~\cite{D2PiMuMu}: {\ensuremath{\Br}$(D^{+}_{s}\rightarrow \pi^+ \eta(\rightarrow \mu^+\mu^-) )$}=(6.8$\pm$2.1)$\times$10$^{-8}$.
Combined with {\ensuremath{\Br}$(D^{+}_{s}\rightarrow \pi^+ \eta )$}=(1.7$\pm$0.09)$\times$10$^{-2}$~\cite{Agashe:2014kda}, this yields 
{\ensuremath{\Br}$(\eta\rightarrow \mu^+\mu^-)$}=(4.0$\pm$1.3)$\times$10$^{-6}$. The weighted average of these two evaluations is
{\ensuremath{\Br}$(\eta\rightarrow \mu^+\mu^-)$}=(4.7$\pm$1.0)$\times$10$^{-6}$. This corresponds to a relative uncertainty of about 20\%. This value can be compared to the PDG average  {\ensuremath{\Br}$(\eta\rightarrow \mu^+\mu^-)$}=(5.8$\pm$0.8)$\times$10$^{-6}$ quoted in the previous sections which is dominated by the SATURNE II  collaboration measurement~\cite{Abegg:1994wx}. The LHCb constraint on  the $\chi_1+\chi_2$ parameter  is shown in Fig.~\ref{Fig:2}. Although the relative error in the branching ratio is 20 \% for LHCb and 14 \% for PDG value, since the branching fraction has parabolic dependence on the $\chi_1+\chi_2$, the LHCb result, which has a smaller central value, leads to a much larger errors in $\chi_1+\chi_2$ for now.  
A simple average of the LHCb and PDG values, $\chi_1(m_\rho)+\chi_2(m_\rho)=-8.4\pm3.4$ or $-30\pm 3.4$, is shown in Figs.~\ref{Fig:2} and~\ref{Fig:3} (the gray bound in Fig.~\ref{Fig:3}). By averaging with the $\pi\to e^+e^-$, we obtain:  
\begin{equation}
\left(\chi_1(m_\rho)+\chi_2(m_\rho)\right)_{\rm average}^{\rm exp}=-13.2\pm 5.0. 
\end{equation}

Now let us discuss possible future LHCb measurements. 
First of all, adding the sample collected in 2012 (2~fb$^{-1}$ at a centre-of-mass energy of 8 TeV), three times more decay candidates would be already available. Therefore, 
with the data already available at LHCb, {\ensuremath{\Br}$(\eta\rightarrow \mu^+\mu^-)$} can
be measured with a relative uncertainty of the order of 13\%. 

The uncertainty above is dominated by the statistical uncertainty on 
{\ensuremath{\Br}$(D^{+}\rightarrow \pi^+ \eta(\rightarrow \mu^+\mu^-) )$} and 
{\ensuremath{\Br}$(D^{+}_{s}\rightarrow \pi^+ \eta(\rightarrow \mu^+\mu^-) )$}. On the other hand, as more data becomes available, the share of uncertainties on 
{\ensuremath{\Br}$(D^{+}\rightarrow \pi^+ \eta )$} and {\ensuremath{\Br}$(D^{+}_{s}\rightarrow \pi^+ \eta )$} (presently known up 
5 to 6\%) will increase. Also, such measurements are in practice normalized to the decays 
$D^{+}_{(s)}\rightarrow \pi^+ \phi(\rightarrow \mu^+\mu^-)$: the value of 
{\ensuremath{\Br}$(D^{+}_{(s)}\rightarrow \pi^+ \eta(\rightarrow \mu^+\mu^-) )$} is derived from the ratio between the yields of 
$D^{+}_{(s)}\rightarrow \pi^+ \eta(\rightarrow \mu^+\mu^-)$ and $D^{+}_{(s)}\rightarrow \pi^+ \phi(\rightarrow \mu^+\mu^-)$ decays 
measured in the sample, combined with the branching fractions {\ensuremath{\Br}$(D^{+}_{(s)}\rightarrow \pi^+ \phi(\rightarrow \mu^+\mu^-) )$}. The latter can be determined with an uncertainty of about 5\% by combining the measured values of {\ensuremath{\Br}$(D^{+}_{(s)}\rightarrow \pi^+ \phi)$} and {\ensuremath{\Br}$(\phi\rightarrow e^+e^-)$}~\cite{Agashe:2014kda} and assuming no difference between the decays 
$\phi\rightarrow \mu^+\mu^-$ and $\phi\rightarrow e^+e^-$.

In the 5 fb$^{-1}$ data sample collected between 2015 and 2018 at a centre-of-mass energy of 13 TeV (LHCb Run II), 
about 3 times more $D$-meson decays should be produced (larger integrated luminosity and twice larger production cross-section).  
Using this data, {\ensuremath{\Br}$(\eta\rightarrow \mu^+\mu^-)$} can therefore be measured with a relative 
statistical uncertainty of about 6\%.  Accounting also for the uncertainty on the normalisation mode and on 
{\ensuremath{\Br}$(D^{+}\rightarrow \pi^+ \eta )$}, we conclude that {\ensuremath{\Br}$(\eta\rightarrow \mu^+\mu^-)$} 
can be measured up to about a 10\% uncertainty with LHCb Run II's data.

The measurement of {\ensuremath{\Br}$(\eta\rightarrow \mu^+\mu^-)$} can be further improved by using the decay {$D^{0}\rightarrow K^-\pi^+ \eta(\rightarrow \mu^+\mu^-)$}.
In.~\cite{Aaij:2015hva}, LHCb measured the branching fraction of the  
$D^{0}\rightarrow K^-\pi^+ \mu^+\mu^-$ process in the dimuon mass range {675 $ < m(\mu^+\mu^-) < $ 875 MeV/$c^2$}. Focussing
on events triggered thanks to the muons in the final state,  roughly 15000 {$D^{0}\rightarrow K^-\pi^+ \mu^+\mu^-$} 
decays can be isolated in the 2 fb$^{-1}$ sample used in~\cite{Aaij:2015hva}, 
where $\sim$5000 background candidates also lies in the signal region of the $m(K^-\pi^+ \mu^+\mu^-)$. In the 
dimuon mass range {538 $ < m(\mu^+\mu^-) < $  558 MeV/$c^2$}, where almost all of {$\eta\rightarrow \mu^+\mu^-$} decays
should be reconstructed, we expect about ten times less background candidates. Using {\ensuremath{\Br}($D^{0}\rightarrow K^-\pi^+ \eta$)}=0.01 and {\ensuremath{\Br}$(\eta\rightarrow \mu^+\mu^-)$}=(5.8$\pm$0.8)$\times$10$^{-6}$ (Table 1) to 
predict {\ensuremath{\Br}$(D^{0}\rightarrow K^-\pi^+ \eta (\rightarrow \mu^+\mu^-))$}$\sim$5.8$\times$10$^{-8}$, we 
rescale the yield of signal events found in {675 $ < m(\mu^+\mu^-)< $ 875 MeV/$c^2$}, where the 
branching fraction is {\ensuremath{\Br}$(D^{0}\rightarrow K^-\pi^+ \mu^+\mu^-)$}=(4.2$\pm$0.4)$\times$10$^{-6}$~\cite{Aaij:2015hva}.
We predict that about 200  $D^{0}\rightarrow K^-\pi^+ \eta (\rightarrow \mu^+\mu^-)$ decays can be found in the 2 fb$^{-1}$ sample 
collected in 2012. This corresponds to a significance of about 7 and suggests 
{\ensuremath{\Br}$(D^{0}\rightarrow K^-\pi^+ \eta (\rightarrow \mu^+\mu^-))$} can be measured with a statistical 
precision of about 15\%. Combined 
with the 10\% uncertainty on {\ensuremath{\Br}$(D^{0}\rightarrow K^-\pi^+ \mu^+\mu^-)$} in {675 $ < m(\mu^+\mu^-)< $ 875 MeV/$c^2$}, 
which would be used as the normalisation 
mode in this case, this would yield a 20\% precise measurement of {\ensuremath{\Br}$(\eta\rightarrow \mu^+\mu^-)$}. This 
also assumes that {\ensuremath{\Br}($D^{0}\rightarrow K^-\pi^+ \eta$)} can be measured precisely at Flavor Factories 
in the coming years. The value assumed above is derived from  {\ensuremath{\Br}$(D^{0}\rightarrow \pi^+\pi^- \eta)$}=(1.09$\pm$0.16)$\times$10$^{-3}$~\cite{Agashe:2014kda} by betting that the branching fraction of $D^{0}\rightarrow K^-\pi^+ \eta$, 
the corresponding Cabibbo favored mode, is an order of magnitude higher. A 20\% precise measurement through this mode
would provide a valuable cross-check of the value obtained through the decays 
{\ensuremath{\Br}$(D^{+}_{(s)}\rightarrow \pi^+ \eta(\rightarrow \mu^+\mu^-) )$}. Combining both measurements, 
an overall 10\% precision on {\ensuremath{\Br}$(\eta\rightarrow \mu^+\mu^-)$} is possible. 
 
Looking further, the Upgrade of LHCb is planed to collect 50 fb$^{-1}$ of data at $\sqrt{s}$=14 TeV. We expect precisions of the order of 8\% for the measurement using $D^{+}_{(s)}\rightarrow \pi^+ \eta(\rightarrow \mu^+\mu^-) $ decays and of 10\% for that using $D^{0}\rightarrow K^-\pi^+ \eta (\rightarrow \mu^+\mu^-)$ decays. The uncertainties  do not decrease much 
due to the fact that we are not accounting for the improvements in the the uncertainties on the normalization modes nor on {\ensuremath{\Br}$(D^{+}_{(s)}\rightarrow \pi^+ \eta )$} and {\ensuremath{\Br}($D^{0}\rightarrow K^-\pi^+ \eta$)}. These errors can be indeed reduced by LHCb or Belle II, BESIII experiments independently, which will further reduce the errors in {\ensuremath{\Br}$(\eta\rightarrow \mu^+\mu^-)$}. 

%%%%%%%%%%%%%%%%%%%%%%%%%%%%%%%%%%%%%%%%
\subsection{Sensitivity of the LHCb experiment to {\ensuremath{\Br}$(\eta^{'}\rightarrow \mu^+\mu^-)$} }
\label{sec:LHCbSensitivityEtaP}
According to the prediction in Table 1, the branching ratio of the  decay  {\ensuremath{\Br}$(\eta^{'}\rightarrow \mu^+\mu^-)$} is 40 
times lower than that of  {\ensuremath{\Br}$(\eta\rightarrow \mu^+\mu^-)$}.  It will therefore be far more difficult to measure. 
It should however be possible using the sample collected with the Upgraded detector,  by measuring
{\ensuremath{\Br}$(D^{+}_{s}\rightarrow \pi^+ \eta^{'}(\rightarrow \mu^+\mu^-) )$}, which is predicted to be as low as 
about 5.5$\times$10$^{-9}$. We evaluate the yield of such decays based on the yield of  $D^{+}_{s}\rightarrow \pi^+ \phi(\rightarrow \mu^+\mu^-)$ 
decays measured in~\cite{D2PiMuMu},  corrected by this ratio~:
\begin{equation}
  \label{eq:Ratio}
   {\cal R} = \frac{{\Br}(D^{+}_{s}\rightarrow \pi^+ \eta^{'} )}{{\Br}(D^{+}_{s}\rightarrow \pi^+ \phi )}.\frac{{\Br}(\eta^{'}\rightarrow \mu^+\mu^- )}{{\Br}(\phi\rightarrow \mu^+\mu^- )},
\end{equation}
where all the branching fractions but {\ensuremath{\Br}$(\eta^{\prime}\rightarrow \mu^+\mu^-)$} are taken from~\cite{Agashe:2014kda}. We also
assume that the Upgraded trigger system will double the signal efficiency. We derive the
background yields that will also contribute to the measurement's uncertainty from the yields observed below the $D^{+}_{s}$ peak of the  
$m(\pi^{+}\mu^{+}\mu^{-})$ distribution as shown in~\cite{D2PiMuMu}. We count only candidates found in the dimuon mass range where the $\phi$ 
was looked for (850-1250 MeV/$c^{2}$), and scale this number by the width ratio between this window and the $\pm$10 MeV/$c^{2}$ window that 
would be used to search for a $\eta^{'}$. For both the signal and background predictions, we also account for the larger integrated luminosity 
and $c-\bar{c}$ production cross-section. With about 250 {$D^{+}_{s}\rightarrow \pi^+ \eta^{'}(\rightarrow \mu^+\mu^-) $} decays and 
about 1000 background events expected in these conditions, a signal significance of 7 is possible and 
{\ensuremath{\Br}$(D^{+}_{s}\rightarrow \pi^+ \eta^{'}(\rightarrow \mu^+\mu^-) )$} is measurable with a precision 
of about 15\%. If so, {\ensuremath{\Br}$(\eta^{'}\rightarrow \mu^+\mu^-)$} will be measured with a similar precision,   
due to the low impact of the uncertainties in 
{\ensuremath{\Br}$(D^{+}_{s}\rightarrow \pi^+ \eta^{'} )$}=(3.94$\pm$0.25)$\times$10$^{-2}$ 
and in
{\ensuremath{\Br}$(D^{+}_{s}\rightarrow \pi^+ \phi )$}=(4.54$\pm$0.16)$\times$10$^{-2}$~\cite{Agashe:2014kda}.
The decay $D^{+}\rightarrow \pi^+ \eta^{'} $ is less competitive due to a lower branching fraction, 
{\ensuremath{\Br}$(D^{+}\rightarrow \pi^+ \eta^{'} )=(4.84\pm 0.31)\times 10^{-3}$}~\cite{Agashe:2014kda}, which is not compensated for by the higher $D^+$ production cross section (3 times higher than that of the $D^+_s$) in proton-proton collisions. Therefore, a 25 \% precise measurement is possible in this channel.

The decay $\eta^{'}\rightarrow \mu^+\mu^-$ can also be studied through the decay $D^{0}\rightarrow K^-\pi^+ \eta^{'} (\rightarrow \mu^+\mu^-)$. 
The same 
computation as in Sect.~\ref{sec:LHCbSensitivityEta}, where 
{\ensuremath{\Br}$(D^{0}\rightarrow K^-\pi^+ \eta^{'} (\rightarrow \mu^+\mu^-))$}$\sim$1.1$\times$10$^{-9}$ 
is substituted for 
{\ensuremath{\Br}$(D^{0}\rightarrow K^-\pi^+ \eta (\rightarrow \mu^+\mu^-))$}, can be 
carried out to conclude that even LHCb's Upgrade can't provide a precise measurement of  such a small branching 
ratio. Another experimental approach is possible: a $tagged$ measurement focussing on 
{$D^{0}\rightarrow K^{-}\pi^{+} \eta^{'}(\rightarrow \mu^{+}\mu^{-})$} decays where the $D^{0}$ originated from $D^{*+}\rightarrow D^{0}\pi^{+}$. 
The mass difference $\Delta m=m(D^{*+})-m(D^{0})$ can be used to reduce the background level by two orders
of magnitude, at the price of about 10 times less signal decays due to a $D^{*+}$ production
cross-section about 2.5 times lower, to {\ensuremath{\Br}$(D^{*+}\rightarrow D^{0}\pi^{+})$}=2$/$3 and to the low
efficiency with which soft pions can be reconstructed and selected. Assuming once more that the Upgrade of the 
experiment will allow to raise the trigger efficiency by a factor 2, we can hope to measure 
{\ensuremath{\Br}$(D^{0}\rightarrow K^-\pi^+ \eta^{'} (\rightarrow \mu^+\mu^-))$} and therefore 
{\ensuremath{\Br}$(\eta^{'}\rightarrow \mu^+\mu^-)$} with a precision of about 30\%. 

We conclude that the Upgraded LHCb experiment can produce the first precise measurement of 
{\ensuremath{\Br}$(\eta^{'}\rightarrow \mu^+\mu^-)$} with an uncertainty lower than  15 \%.

\section{Conclusions}
The $\eta\rightarrow \mu^+\mu^-$ process is known to provide crucial information for \corrn{various low energy phenomena, including the light-by-light contribution to the muon anomalous magnetic moment}. The theoretical model which is often used to estimate the light-by-light contribution predicts rather well the branching ratio of the $\eta\rightarrow \mu^+\mu^-$ process. \corrn{However,} the latest experimental results on the $\pi^0\to e^+e^-$ branching ratio by the KTeV collaboration shows a tension with the prediction from the same model. Since $Br(\eta\rightarrow \mu^+\mu^-)$ average is dominated by a single measurement, a confirmation of this result is desirable (see~\cite{Nyffeler:2016gnb, Masjuan:2015cjl} for recent discussions). 

In this \corr{article}, we propose a novel approach to determine the branching ratios of $\eta\rightarrow \mu^+\mu^-$ decays using the high static LHCb data of charmed meson decays. 
Based on the $D^{+}_{(s)}\rightarrow \pi^+ \eta (\rightarrow \mu^+\mu^-)$ decays already observed by
LHCb in the 1 $fb^{-1}$ sample collected in 2011,
we find that this branching fraction could be measured with a
precision of 20 \%. This should be compared with the 13\% uncertainty of the present world average:
$Br(\eta\rightarrow \mu^+\mu^-)=(5.8\pm 0.8)\times 10^{-6}$. 
Using the LHCb Run II data ($5fb^{-1}$, $\sqrt{s}$=13 TeV), we estimate that \corrn{the statistical uncertainties} can be reduced down to $\sim 10$ \% level. The {$D^{0}\rightarrow K^-\pi^+ \eta (\rightarrow \mu^+\mu^-)$} decay chain can be also used to further improve the precision. 
We estimate that using 2$fb^{-1}$ from Run I data,
$Br(\eta\rightarrow \mu^+\mu^-)$ can be determined at the 20 \% level from this channel. \corrn{This assumes} 
that the branching fraction of the first stage of the decay chain,
$Br(D^{0}\rightarrow K^-\pi^+\eta)$, can be determined at the few
percent level using non leptonic decays of the $\eta$,  which
seems in the reach of
Belle II and BESIII.
With this additional information, the knowledge of {\ensuremath{\Br}$(\eta\rightarrow \mu^+\mu^-)$} 
would be improved
with respect to the present situation, both in precision
(by about  a factor two)  and in robustness, due to the existence of
two new independent analyses. 

 With the LHCb upgrade, another factor 10 in statistics will be gained and 
will make this extremely rare channel easier to observe. By then, the uncertainty on the $Br(\eta\rightarrow \mu^+\mu^-)$ will be  limited by the knowledge of the normalization and the reference channels. The measurements of these channels by LHCb as well as the $e^+e^-$ colliders such as Belle II and BESIII will become essential to obtain the required high precision measurement.  

We have also studied a possible first observation  of the decay $\eta^{\prime}\rightarrow \mu^+\mu^-$ which the theoretical model mentioned above predicts to be 40 times suppressed \corrn{with respect to} the  $\eta$ decay. 
\corrn{With the LHCb upgrade}, this branching ratio could be measured with an uncertainty below 15 \% via the 
$D_{(s)}^+\rightarrow \pi^+ \eta^{\prime} (\rightarrow \mu^+\mu^-)$ channels. 
With an improved trigger efficiency, the $D^{0}\rightarrow K^-\pi^+ \eta^{\prime} (\rightarrow \mu^+\mu^-)$ channel could also be used to confirm an early observation in the previous mode, which would challenge the theoretical prediction.

\section*{Acknowledgement}
\corrn{We would like to acknowledge B. Moussallam for collaborations at the beginning of this project and many advices. 
N.~T.~H. thanks LAL and IPNO for the hospitality and the financial support  during her stay in Orsay.  }

 \end{document}